\documentclass[twocolumn,preprintnumbers,amsmath,amssymb]{revtex4}

\usepackage{graphicx}% Include figure files
\usepackage{dcolumn}% Align table columns on decimal point
\usepackage{bm}% bold math
\usepackage{amssymb}
\usepackage{amsmath}
\usepackage{enumerate}
\usepackage{wrapfig,epsfig}
\usepackage{graphics}
\usepackage{shadow}
\usepackage[T1]{fontenc}
\usepackage{color}
\usepackage{array}
\usepackage{rotating}

\begin{document}

\title{A microfluidic device for investigating crystal nucleation kinetics}
\author{Philippe Laval}
\email{philippe.laval-exterieur@eu.rhodia.com}
\affiliation{LOF, unité mixte Rhodia--CNRS--Bordeaux 1, 178 avenue du Docteur Schweitzer, F--33608  Pessac cedex -- FRANCE}
\author{Jean-Baptiste Salmon}
\affiliation{LOF, unité mixte Rhodia--CNRS--Bordeaux 1, 178 avenue du Docteur Schweitzer, F--33608  Pessac cedex -- FRANCE}
\author{Mathieu Joanicot}
\affiliation{LOF, unité mixte Rhodia--CNRS--Bordeaux 1, 178 avenue du Docteur Schweitzer, F--33608  Pessac cedex -- FRANCE}
\date{\today}
\begin{abstract} 
We have developed an original setup using microfluidic tools allowing one to produce continuously monodisperse microreactors ($\approx 100$~nL), and to control their temperatures as they flow in the microdevice. With a specific microchannels geometry, we are able to apply large temperature quenches to droplets containing a KNO$_3$ solution (up to 50$^{\circ}$C in 10~s), and then to follow nucleation kinetics at high supersaturations. By measuring the probability of crystal presence in the droplets as a function of time, we estimate the nucleation rate for different supersaturations, and confront our results to the classical nucleation theory.
\end{abstract}
\maketitle

\section{Introduction}
Nucleation kinetics plays a fundamental role in crystallization processes. In particular, it is responsible for the final properties of the crystals:  size, size distribution and polymorphs. Therefore, the control of nucleation kinetics is a real challenge in many scientific fields like proteomics, chemical engineering, pharmacology\dots \cite{LACMANN1999,RODRIGUEZHORNED1999,LUFT2001,COUPLAND2002,VEESLER2003}.
The nucleation rate $J$, the number of nuclei produced by unit of time and volume, is the relevant parameter characterizing nucleation kinetics. In the classical nucleation theory, $J$ is given by:
\begin{equation}
J = AS\exp{-\frac{B}{(\log S)^{2}}}\label{J}\,,
\end{equation}
where $A$ is a kinetic factor, $S$ the supersaturation of the solution, and $B$ depends on the interfacial tension $\gamma$ between the nuclei and the supersaturated solution \cite{ZETTLEMOYER1969,MULLIN2001,KASHCHIEV2003A}. $S$ corresponds to the ratio of the activities of the solution in the supersaturated state and at equilibrium. The above expression of $J$ holds for both homogeneous and heterogeneous nucleation, only the values of $A$ and $B$ differ. The classical nucleation theory accounts well for many experimental facts, and some experiments have yielded values of $A$ and $B$ which are close to theoretical predictions for homogeneous nucleation in the case of precipitation \cite{NIELSEN1961,NIELSEN1969}, solidification (polypropylene, n-octadecane\dots) \cite{VONNEGUT1948,TURNBULL1961}. However, homogeneous nucleation has not yet been observed to our knowledge for crystallization of solutes, due to the unavoidable presence of impurities. Moreover, it seems nowadays that the classical nucleation theory is too much simplified to describe the nucleation process in solution. For instance, it does not consider the desolvatation energy, and several authors propose new mechanisms involving several steps during the nucleation \cite{WOLDE1997,ANDERSON2002,VEKILOV2005,KASHCHIEV2005}. 

One classical method to estimate the nucleation rate $J$ consists of the determination of the induction time $t_{i}$ which is the time elapsing between the creation of the supersaturation (e.g. obtained after a temperature quench) and the first appearance of particles by some  detection device. One has:
\begin{equation} 
t_{i} = t_{n} + t_{g}\,,
\end{equation}
where $t_{n}$ is the time required for the nucleation, and $t_{g}$ the growth time of the nucleus to the detectable dimension. However, it is difficult to determine $J$ from $t_{i}$ measurements since it depends on the sensitivity of the detection device (growth kinetics of the crystals has to be known) \cite{KASHCHIEV2003A}. Moreover, in this type of experiments, usually carried out in macroscopic volumes, impurities play a significant role, a rapid temperature quench is difficult to achieve, and mixing can have an effect \cite{MULLIN1962}.

Actually, the most reliable way to measure crystal nucleation kinetics is the \textit{droplets method} \cite{VONNEGUT1948,TURNBULL1952,POUND1952,WHITE1959,MELIA1964,WOOD1970}. The principle is to divide a volume of solution in a large number of small independant reactors, for example droplets in suspension in an inert oil. In this case, one may observe homogeneous nucleation if the number of droplets is larger than the impurities initially present in the solution. After supersaturation is reached (generally by cooling), one measures the fraction of droplets which contain a crystal as a function of time.
When the droplets volume $V$ is small enough, the nucleation time $t_{n}$ is large compared to $t_{g}$, and so only mononuclear nucleation occurs \cite{KASHCHIEV1994}. In that case, nucleation being a stochastic process, the probability $P$ that a droplet contains a crystal evolves as:
\begin{equation}
P(t) = 1 - \exp(-\omega t)\,,\label{proba}
\end{equation}
when the solution is instantaneously supersaturated at $t = 0$ \cite{ZETTLEMOYER1969}. $\omega =JV$ corresponds to the nucleation frequency, namely the number of nuclei formed per unit of time. With this method, the simple observation of a large number of drops enables one to reach an accurate statistics on $P$ and so to determine the nucleation rate.

Nevertheless, the droplets method raises some experimental difficulties. Indeed, droplets are generally produced by emulsification of the solution in oil using surfactants, and are never perfectly monodisperse. Thus, it is necessary to measure their polydispersity to determine $P(t)$, and surfactants may induce heterogeneous nucleation \cite{DAVEY1997}. It has also been shown that nucleation events in drops may not be independant in concentrated emulsions \cite{HERHOLD1999}. Moreover, detection of the fraction of crystallized drops is often indirect, and rapid temperature quenches are difficult to achieve.

In this paper, we present a new microfluidic tool suitable for kinetic studies of crystal nucleation. This device, based on the droplets method, allows us to overcome most of the difficulties described above. Indeed, it enables us to produce monodisperse droplets of solution {\it without surfactants}, to apply {\it very rapid} temperature quenches, and to measure {\it directly and continuously} the proportion of crystallized droplets.
In a few words, in the device, monodisperse droplets of the solution are continuously produced in an oil phase and convected along a microfluidic channel. During the flow, the droplets are quickly cooled down (typically 10~s) to a temperature below their solubility temperature so that nucleation can occur. In the microchannel, the time of flow $t$ of the droplets is related to the distance $d$ they cover along the channel by $t=d/U$, with $U$ the droplets velocity. As a consequence, we can perform stationary measurement of a kinetic process along the flow \cite{JOANICOT2005}. Thus, to determine the nucleation rate, we just need to measure the probability $P$ at different positions along the channel on a large number of droplets to obtain a satisfactory statistics.
In the present work, we use this microfluidic tool to measure the nucleation rate of a solution of potassium nitrate (KNO$_{3}$) in water for different supersaturations. We show that the device is suitable for rapid measurements of $J$ as a function of $S$, and, using the classical nucleation theory, to estimate physical properties such as the interfacial tension $\gamma$.

\begin{figure}[htbp]
\begin{center}
\scalebox{1.0}{\includegraphics{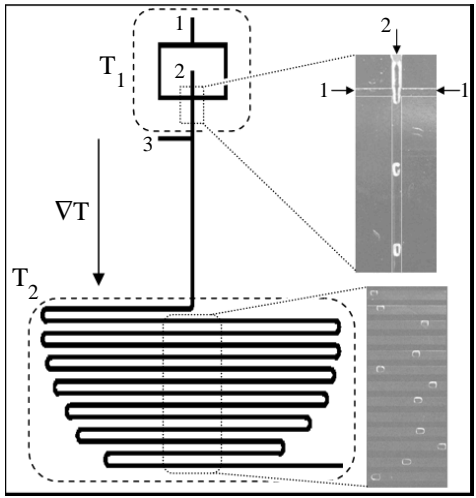}}
\caption{\small Design of the microdevice. The oil and aqueous phases are injected respectively in inlets 1 and 2. Oil can also be injected in inlet 3 to increase the velocity of the droplets. The top-right image shows the formation of the droplets. The two dashed areas are temperature controlled at $T_{1}$ and $T_{2}$. The dotted area in the middle of the serpentine corresponds to the observation zone through a stereo microscope.}
\label{design}
\end{center}
\end{figure}

\section{Experimental section}

\subsection{Microfluidic experiments}
Microfluidics refers to systems used to manipulate liquids in channels with typical dimension of 100~$\mu$m (see \cite{STONE2004,SQUIRES2005} and references therein). In specific channels geometries, a two-phases flow can produce monodisperse droplets \cite{ANNA2003}.
The design of the microdevice is shown on Fig.~\ref{design}. It is composed of three inlets and one outlet. In inlets 1 and 3, viscous silicone oil (Rhodorsil 500~cSt) is injected at constant flow rate ranging between 1 and 4~mL~hr$^{-1}$. The aqueous solution used for the study is previously prepared in a beaker. It was made of 83.6~g of KNO$_{3}$ (Normapur Merck) added to 100~g of deionized water. The solubility temperature $T_{s}$ of this solution is 50$^{\circ}$C \cite{MULLIN2001}. It is injected in inlet 2 at flow rate $Q$ of about $500~\mu$L~hr$^{-1}$. The liquids are injected with syringe pumps (PHD 2000 infusion Harvard Apparatus). At the intersection between the oil and the aqueous streams, monodisperse droplets are continuously produced (see up right insert on Fig.~\ref{design}). It is important to note that to avoid nucleation at the surfaces of the droplets we choose silicone oil which does not wet KNO$_{3}$ crystals.
The droplets volume $V$ is controlled by the ratio of oil (from inlet 1) to aqueous phase flow rates and measured by $V=Q/f$, where $f$ is the droplets production frequency (close to 0.3 droplet per second in our device). With the geometry of the microchannels used in these experiments, the droplets volume could be tuned from 50 to 200~nL. The monodispersity of the droplets was determined by image analyses measuring the time $t_{d}$ between the production of two successive droplets. Since $Q=V/t_{d}$, and $Q$ is constant, variations of $t_{d}$ are directly related to the variations of $V$. We find a {\it volume} polydispersity of a few percents.
When the flow rates ratio is changed to tune the volume of the droplets, the total flow rate of liquids can be modified and as a consequence, the velocity of the droplets is also changed. To balance these variations of the total flow rate, and keep constant the velocity of the droplets, silicone oil is injected through inlet 3.

\subsection{Microfabrication}
The device is fabricated in poly(dimethylsiloxane) (PDMS) by using soft-lithographic techniques \cite{MCDONALD2002}. PDMS (Silicone Elastomer Base, Sylgard 184; Dow Corning) was molded on master fabricated on a silicon wafer (3-Inch-Si-Wafer; Siegert Consulting e.k.) using a negative photoresist (SU-8 2100; MicroChem). Microchannels of 500~$\mu$m high were used. To make molds of such heights, we spin successively two 250~$\mu$m thick SU-8 layers on the wafer. After each spincoating process, the wafer is soft-baked for 10~min at 65$^{\circ}$C and for 60~min at 95$^{\circ}$C. Photolithography was used to define negative images of the microfluidic channels. Finally the wafer was hard-baked for 25~min at 95$^{\circ}$C and developed (SU-8 Developer; MicroChem).

In order to have all channel walls made of the same material, the PDMS-molded channels were sealed with a silicon wafer previously covered by a thin layer of cured PDMS of about 50~$\mu$m. To achieve this sealing, a mixture of $80\%$~wt. in PDMS and $20\%$~wt. in curing agent (Curing Agent Silicone Elastomer, Sylgard 184; Dow Corning) was molded on the SU-8 master described above at 65$^{\circ}$C for 25~min. At the same time, a mixture of $95\%$~wt. in PDMS and $5\%$~wt. in curing agent was spun on the wafer to produce a thin layer and was baked at 65$^{\circ}$C for 40~min \cite{UNGER2000}. The device is then peeled off the mold and holes for the inlets and outlets (1/32~in. o.d.) were punched into the material. Eventually, the device was placed with the wafer covered by the layer of PDMS in a UV ozone apparatus (UVO Cleaner, Model 144AX; Jelight) and exposed to irradiation for 2~min, before sealing at 65$^{\circ}$C for at least one day.

\subsection{Observation}
\begin{figure}[htbp]
\begin{center}
\scalebox{1.0}{\includegraphics{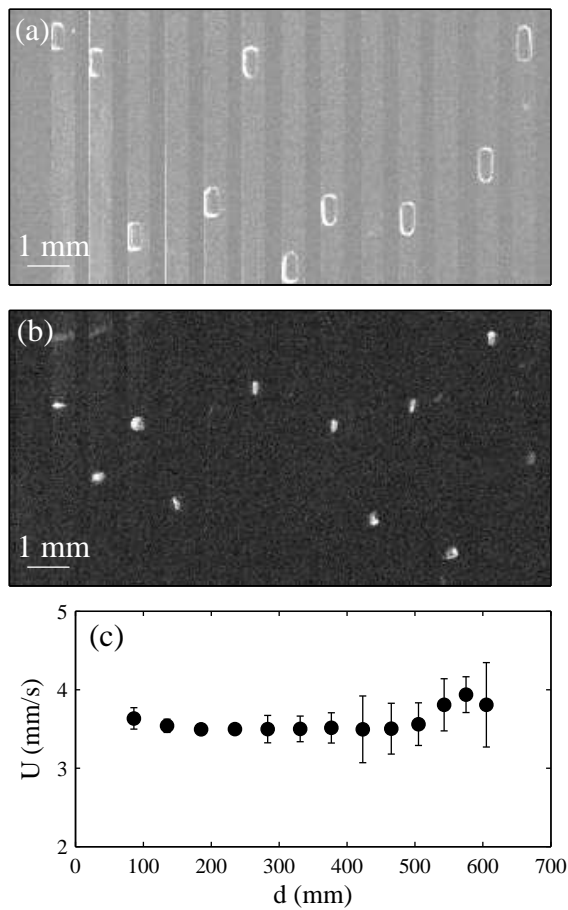}}
\caption{\small (a) Typical image of a movie of the observation area in the serpentine. Such images enable us to measure the velocity of the droplets along the channel. (b) Image of the same area obtained under crossed polarizers. Only birefringent objects such as crystals of KNO$_{3}$ can be observed. (c) Velocity $U$ of the droplets as a function of their position along the microchannel.}
\label{imacanaux}
\end{center}
\end{figure}

To study the nucleation kinetics in the microdevice, we observe all the dotted area on Fig.~\ref{design} with a stereo microscope (SZX12; Olympus) and a CCD camera (C4742-95; Hamamatsu). Thanks to the serpentine shape of the channel, the observation of this area allows us to access simultaneously different times of the kinetics. A typical image of this area is shown on Fig.~\ref{imacanaux}. Such images are used to measure the velocity $U$ of the droplets along the channel. More precisely, to determine $U$, we make a movie of the observation area during the time required for the passage of about 30 droplets through each section. An image processing program, searching the maximum of auto-correlation between two successive images of the movie, measures automatically the velocity of each droplet. Then, the average of the velocities $U$ is calculated for each section of the serpentine, that is to say, for different distances along the microchannel. These measurements, which show on Fig.~\ref{imacanaux}c a slight variation of the velocity (less than 10$\%$) probably due to a small inhomogeneity of the channel height, enable us to achieve an accurate estimation of the time of flow of the droplets.

The detection of KNO$_{3}$ crystals formed in the droplets is based on their birefringent properties. To measure the proportion of droplets that contain a crystal as a function of time, we make another movie of the observation area but under crossed polarizers and during the time $t_{m}$ corresponding to the passage of about 150 droplets. Because of their birefringence, we only see crystals (see Fig.~\ref{imacanaux}b). An image processing program counts automatically the numbers $N$ of crystals that go through each serpentine section during $t_{m}$. From the droplets production frequency $f$, we then determine the total number $N_{0}$ of droplets that go through each section during the movie. 
For an experiment in which $N$ events are observed in $N_{0}$ trials, the probability to observe an event is $P=N/N_{0}$. The error on the determination of $P$ is given by the standard deviation of the binomial distribution function of $P$ in the case of independant events. Experimentally, $P$ is determined from a statistics on about 150 droplets, the results being the same as for experiments carried out on 500 droplets.

%which appear to be satisfactory since we obtain the same results as for experiments carried out on 500 droplets. 

\subsection{Temperature control}
To avoid any crystallization before the droplets formation, the syringe containing the solution and the corresponding tubing are heated at about 60$^{\circ}$C with two flexible heaters (Minco) and, as shown on Fig.~\ref{design}, the inlets area is heated at a temperature $T_{1}$ also with a heater (Minco) placed on the back of the silicon wafer. Each heater is controlled with a temperature controller (Minco). The serpentine area is kept at temperature $T_{2}$ using a thermostated support placed underneath. The temperature of the support is regulated with a water circulation and a cryostat (F25; Julabo).

\begin{figure}[htbp]
\begin{center}
\scalebox{1.0}{\includegraphics{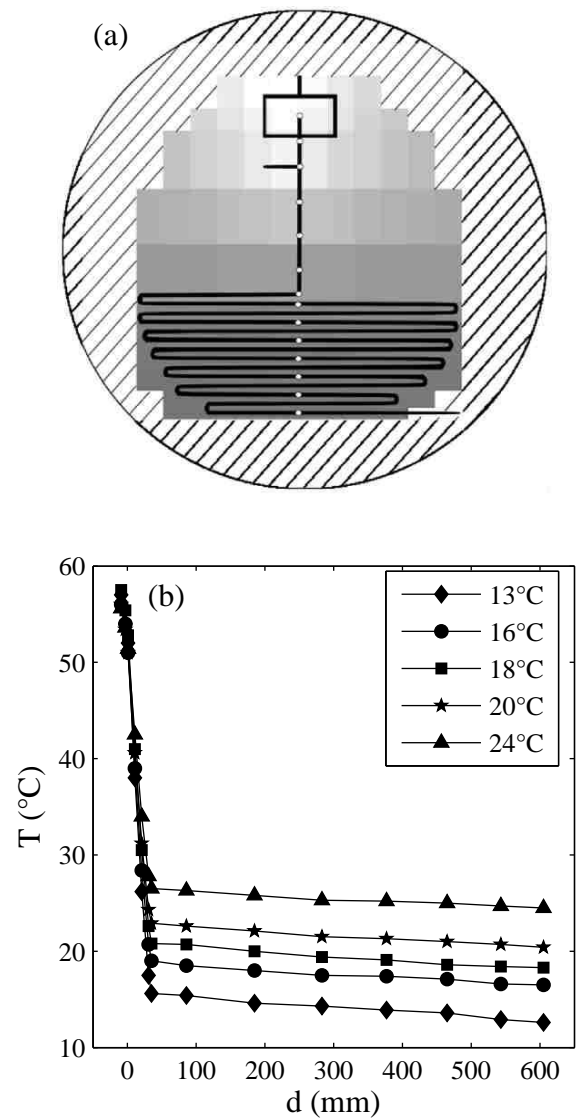}}
\caption{\small (a) Temperature profile of the device.The gray scale is linear with the temperature. The hot area is in white and the cold one is in dark grey. The hatched disk corresponds to the silicon wafer. (b) Temperature of the droplets flowing along the channel for different temperatures $T_{2}$ of the cooled area. The positions along the channel where the temperatures are measured, are marked by white spots on image (a).}
\label{temperatures}
\end{center}
\end{figure}

For the temperature measurements in the device, thin thermocouples (type K, 0.5~mm o.d.; Thermocoax) are inserted through the PDMS layer down to the silicon wafer. These thermocouples enable us to measure the temperature close to the microchannel. Figure~\ref{temperatures}a, displaying a typical temperature profile, shows the local heating of the inlets area and the homogeneous temperature of the serpentine one. We have measured the temperatures at different positions along the channel, marked by the white spots on Fig.~\ref{temperatures}a, for five different minimal temperatures $T_{2}$ of the serpentine area. These temperature profiles are plotted on Fig.~\ref{temperatures}b. They demonstrate that we can apply a large temperature gradient between the inlets area and the beginning of the serpentine (typically up to 50$^{\circ}$C, from 60 to 10$^{\circ}$C), maintaining the serpentine area at a temperature almost constant, with maximal variations of 1 to 2$^{\circ}$C.

The temperature of the liquid flowing along the channel has been measured with thermocouples (type K, $0.5~$mm o.d.; Thermocoax) drove into the microchannel. We have observed that, anywhere along the channel, the liquid is at the same temperature as the wafer. This is mainly due to the small velocity of the droplets (typically $4~$mm~s$^{-1}$), the small height of the microchannel, the forced convection of liquid in the droplets \cite{SONG2003A}, and the high thermal conductivity of the silicon wafer. As a consequence, as soon as the droplets are produced, they are quenched down to the serpentine temperature (typically 50$^\circ$C in about 10~s) which allows us to access kinetics with characteristic times ranging from 10~s to the total residence time of about 150~s (for $U\approx4$~mm~s$^{-1}$).

\section{Results and discussion}

\begin{figure}[htbp]
\begin{center}
\scalebox{1.0}{\includegraphics{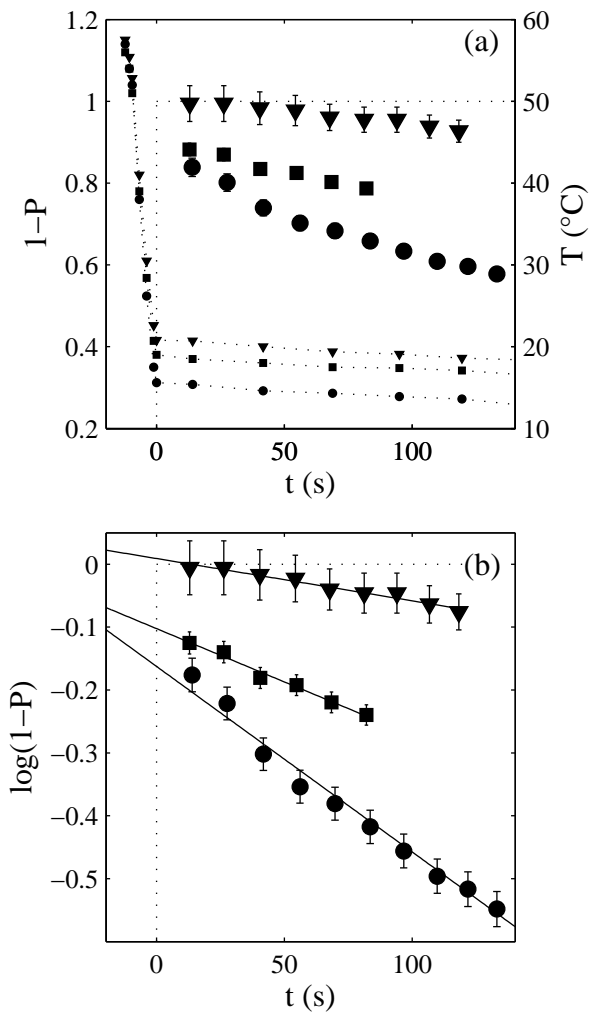}}
\caption{\small (a) $1-P$ and the droplets temperature (dotted plots) as a function of time. $P$ is the probability to have a crystal in a droplet. (b) $\log(1-P)$ as a function of time. These measurements have been obtained with droplets of 100 nL and for three different temperatures of the cooled area (bullets: 13$^{\circ}$C, squares: 16$^{\circ}$C and triangles: 20$^{\circ}$C). The time origin corresponds to the entrance of the droplets in the serpentine.}
\label{Probability}
\end{center}
\end{figure}

We now discuss the results obtained for the KNO$_3$ solution whose temperature solubility is 50$^\circ$C (see experimental section). The droplets are formed at 60$^\circ$C and cooled down to different temperatures $T_{2}$. For $T_{2}>25^\circ$C, no nucleation events occur when droplets flow along the channel and for $T_{2}<10^\circ$C, crystals systematically appear during the temperature quench. Figure~\ref{Probability}a presents the temporal evolution of $1-P$ for three intermediate temperatures $T_{2}$, where $P$ is the probability  that a drop contains a crystal.
The origin of the time axis is set at entrance of the serpentine. As expected, $1-P$ decreases with time and the kinetics of crystal nucleation becomes faster as $T_{2}$ decreases (Eq.~(\ref{proba})). Note that, at the investigated range of temperatures, growth kinetics (experimentally measured of the order of 1~s) is very fast compared to the nucleation kinetics (typically 50~s) so we can assume that the time at which a crystal is first observed corresponds to the nucleation event. 
As shown by Eq.~(\ref{proba}), $1-P$ is expected to follow an exponential decay characterized by the nucleation frequency $\omega$. However, the results displayed in Fig.~\ref{Probability}b, can not be accounted by simple exponential behaviours but rather by $1-P=\exp(-\omega t+\epsilon)$. Such offsets at $t=0$~s have already been reported and can be explained by the presence of impurities in some of the droplets \cite{WHITE1959,HERHOLD1999,GALKIN1999}. Indeed, they may provoke heterogeneous nucleation with faster kinetics. Therefore we observe a fraction of droplets which contain crystals before the first observed channel ($\approx 25$~s in our device). Surprisingly, slightly positive offsets, corresponding to delay times for the kinetics, have been systematically observed at temperatures higher than 20$^\circ$C, where probabilities $P$ do not exceed $5\%$ at $t=150$~s. This seems to indicate that there is no fast kinetics induced by impurities as above. In this case, the observed delay times may be due to slight variations of the temperature along the serpentine which can play an important role at low probabilities.

\begin{figure}[htbp]
\begin{center}
\scalebox{1.0}{\includegraphics{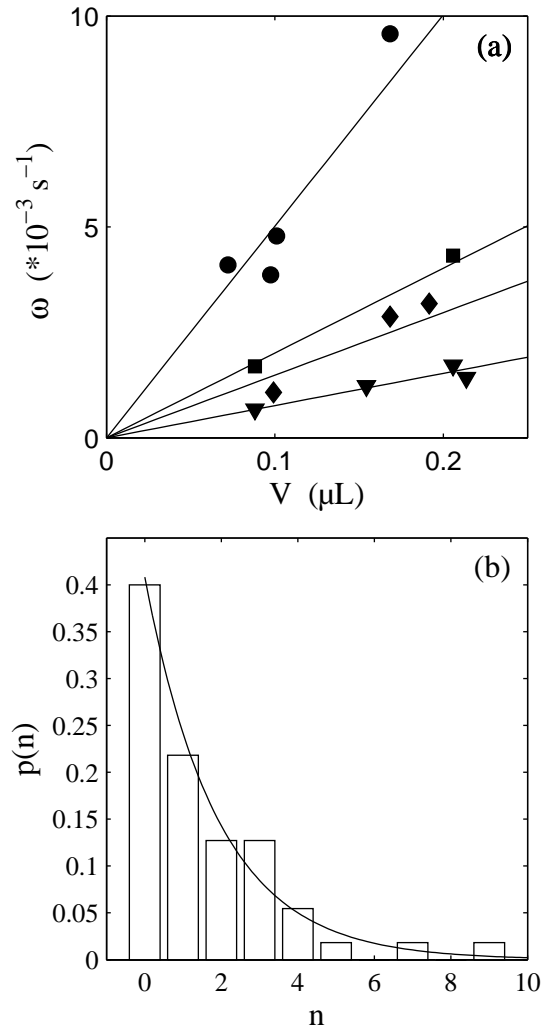}}
\caption{\small (a) Nucleation frequency $\omega$ as a function of droplet volume $V$ for different temperatures of the cooled area (bullets: 13$^{\circ}$C, squares: 16$^{\circ}$C, diamonds: 18$^{\circ}$C and triangles: 20$^{\circ}$C). The solid lines are the linear fits of the plots for the different temperatures. (b) Probability $p$ to have $n$ droplets without crystal between two droplets containing a crystal, for $P~=~0.41$. The bars correspond to the experimental data obtained for 200 droplets and the solid line is the geometrical law $p(n)=P(1-P)^{n}$.}
\label{frequenceNucl}
\end{center}
\end{figure}

Despite the observed offsets, the nucleation frequencies $\omega$ are estimated by fitting the data according to $1-P=\exp(-\omega t+\epsilon)$. $\omega$ are almost constant over the investigated range of time. Figure~\ref{frequenceNucl}a presents the nucleation frequency $\omega$ as a function of the droplets volume $V$ for different temperatures $T_{2}$ (in all these experiments, the velocity of the droplets is almost the same). Our results suggest that $\omega$ is proportional to $V$ at a given temperature and therefore, that nucleation rates can be estimated by $J=\omega /V$. Even if the range of volume variation is not very large, these results indicate that $\omega$ is proportional to the number of nucleation sites inside the droplets and that nucleation occurs in the droplet volume.

In the droplets method, it is important to verify that the presence of a crystal in a droplet does not influence the nucleation in others \cite{HERHOLD1999} (e.g. contamination of the channel in our device). When the nucleation events are independent, the probability $p$ to observe $n$ droplets without crystal between two droplets containing a crystal, must follow the geometrical law
\begin{equation}
 p(n)=P(1-P)^{n}\,.
\end{equation}
We have measured the probability $p(n)$ for different probabilities $P$ in several experiments. On Fig.~\ref{frequenceNucl}b, we compare the experimental results obtained at $16^{\circ}$C for $P=0.41$ with the corresponding geometrical law. Good agreements between the two plots are systematically observed, especially when the statistics is carried out on a large number of droplets. We conclude that the presence of a crystal in a droplet does not induce nucleation in others, and our device is therefore well-suited to follow independent nucleation events.

By changing the temperature $T_{2}$, we could measure the nucleation rate $J$ for different supersaturations $S$. To calculate $S$, we assumed that $S\approx(C_{in}/C_{eq})^{2}$, where $C_{in}$ and $C_{eq}$ are respectively the initial and equilibrium  concentrations of KNO$_{3}$ (despite numerous literature researches, no experimental data could be found concerning the ions activity at the investigated metastable states).

\begin{figure}[htbp]
\begin{center}
\scalebox{1.0}{\includegraphics{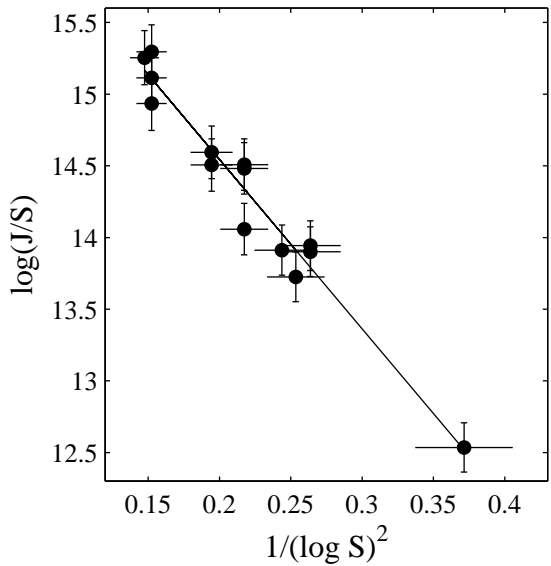}}
\caption{\small (Plot of $\log J/S$ as a function of $1/(\log S)^{2}$. $J$ is the nucleation rate (m$^{-3}$~s$^{-1}$), $S$ the supersaturation. The solid line is the linear regression of the data.}
\label{Jfct2S}
\end{center}
\end{figure}

In the performed experiments, $S$ has been varied in the range of 0 to 13. Below $S=5$, no crystals nucleate within the investigated time (150~s) and above $S=13$, crystals systematically appear during the temperature quench (10~s). Figure~\ref{Jfct2S} presents $\log J/S$ as a function of $1/(\log S)^{2}$. The error bars take into account the small temperature variations along the serpentine.
Within the experimental uncertainties, Eq.~(\ref{J}) well fits our data with $A\approx3.10^{7}~$m$^{-3}$~s$^{-1}$ and $B\approx 12$. In this case, we have supposed that both $A$ and $B$ do not depend on temperature. In accordance with the classical nucleation theory and in the case of spherical nuclei, $B$ is given by
\begin{equation}
B=\frac{16\pi\gamma^{3}\nu^{2}}{3(kT)^{3}}\,,\label{B}
\end{equation}
where $\gamma$ is the interfacial tension between the nucleus and the solution, $\nu$ the molecular volume in the crystal, and $k$ the Boltzman constant. Supposing that $\gamma$ is proportional to $T$ (i.e. $B$ does not depend on $T$), we found  $\gamma/T=6.6\,10^{-2}~$mN~m$^{-1}$~K$^{-1}$ with $\nu=8\,10^{-29}$~m$^{-3}$ \cite{HANDBOOK2004}. In the investigated range of $T$, $\gamma \approx 19\pm 1$~mN~m$^{-1}$. Note that if we suppose $\gamma$ does not depend on $T$, Eq.~(\ref{J}) and Eq.~(\ref{B}) give also a correct fit of the data with $\gamma \approx 20$~mN~m$^{-1}$.

In the case of homogeneous nucleation, theoretical estimations of $A$ are of the order of $10^{26}$--$10^{30}~$m$^{-3}$~s$^{-1}$. The $A$ we measured ($A\approx3\,10^{7}~$m$^{-3}$~s$^{-1}$) suggests we observe heterogeneous nucleation. Moreover, over several series of experiments (corresponding to observation of about 6000--8000 droplets), we found a small dispersion of the $\gamma$ values (16--26~mN~m$^{-1}$) and values of $A$ ranging from $3\,10^{7}$ to $10^{10}$ ~m$^{-3}$~s$^{-1}$ also suggesting a heterogeneous mechanism. Furthermore, since the typical concentration of impurities is about $10^6$--$10^8$~mL$^{-1}$ according to \cite{ZETTLEMOYER1969}, each droplet of 100~nL contains about $10^2$ to $10^4$ impurities susceptible for acting for heterogeneous nucleation.

Nevertheless, it is difficult to tell if we observed homogeneous or heterogeneous nucleation just from comparisons with the classical nucleation theory. Indeed, up to now, no experimental values of $A$ as large as $10^{26}$--$10^{30}~$m$^{-3}$~s$^{-1}$ have been measured for the crystallization of solutes (except for the case of precipitation where high $S$ can be reached), and the classical nucleation theory may not take into account the entire complexity of solute crystallization \cite{MULLIN2001,ZETTLEMOYER1969,KASHCHIEV2003A,WOLDE1997,ANDERSON2002}. For exemple, according to the classical nucleation theory, the kinetic factor $A$ is proportional to the concentration of nucleation sites in the solution \cite{KASHCHIEV2003A}. As a consequence, we can write:
\begin{equation}
\frac{A_{\text{HO}}}{A_{\text{HE}}}=\alpha\frac{C_{0}}{C_{i}},\label{Ratio}
\end{equation}
where $A_{\text{HO}}$ and $A_{\text{HE}}$ are the kinetic factors in the case of homogeneous and heterogeneous nucleation respectively, $C_{0}$ is the molecule concentration in the solution, $C_{i}$ is the concentration of heterogeneous nucleation sites and $\alpha$ is a factor of the order of 1 \cite{KASHCHIEV2003A}. With $A_{\text{HO}}=10^{26}$~m$^{-3}$~s$^{-1}$, $C_{0}=10^{28}$~m$^{-3}$ and $A_{\text{HE}}=10^{7}$~m$^{-3}$~s$^{-1}$ in our case, we have $C_{i}\approx 10^{9}$~m$^{-3}$. Since the droplets volume is 100~nL, this results would suggest that there are less than one heterogeneous nucleation site per droplet, which is contradictory with our previous conclusion.

To bring new experimental insights, one should increase the supersaturation \cite{NIELSEN1961,MULLIN2001} in the solutes crystallization experiments. In our device, all crystals nucleate during the temperature quench at $S \geq 13$ for droplets volume of 100~nL. To reach higher supersaturations and measure the corresponding kinetics, we should decrease significantly the droplets volume down to a few nL, such tiny droplets would allow us to access homogeneous nucleation \cite{ZETTLEMOYER1969}.

\section{Conclusion}
To conclude, we have performed original microfluidic experiments suitable for nucleation kinetics measurements. Indeed, our microdevice enables us to overcome the difficulties encountered with the classical droplets method. More precisely, we have managed with our device to apply rapid and large temperature quenches to monodisperse microreactors, to transport solid particles in droplets, and to monitor continuously with direct observations, the fraction of crystallized droplets as a function of time. Such experiments yield rapid statistical measurements of the nucleation kinetics and therefore give access to value of nucleation rates with a low liquid consumption (200 droplets $\approx 20~\mu$L). 

We also believe our system is well-adapted to follow kinetics of any temperature dependent processes.  With only a few enhancements of the experimental device (automation of several liquids injection to vary solvent and solutes concentration), high-throughput screening of crystallization conditions could be carried out.

\acknowledgments{We gratefully thanks A. Ajdari, G. Cristobal, M. Kohl, J. Leng, C. Mousset, and E. Plasari for fruitful discussions. We also acknowledge {\it Région Aquitaine} for funding and support, and the {\it Atelier Mécanique} of the CRPP for their technical help. }

\end{document}